\documentstyle{article}
\begin{document}
\begin{center}

{\bf
Spin and its evolution for isolated neutron stars and X-ray binaries:
the determination of the 'Diffusion coefficients' and SPINDOWN THEOREM.
}

\vskip 0.5cm

        V.M.Lipunov $^{1,2}$ and S.B.Popov $^{1}$\\
 $^{1}$ Department of Physics, Moscow University, 119899, Moscow, Russia,\\
 $^{2}$ Sternberg Astronomical Institute, 119899,
 Universitetsky pr., 13, Moscow, Russia\\
\end{center}

\vskip 0.5cm

\begin{abstract}

\vskip 0.1cm

 In this work we give detail consideration of the possible scenario
of evolution of isolated neutron stars (INSs) and determine some
characteristics of X-ray pulsars from their spin period evolution.

\vskip 0.2cm

 The new points of our consideration are:

\vskip 0.1cm

\noindent
--we give additional arguments for the short time scale of the Ejector
stage ( $ \approx 10^7-10^8 $ yrs ).

\vskip 0.2cm

\noindent
--we proposed specific SPINDOWN THEOREM and give some arguments for its
validity. Discovery of accreting INSs will means that the
SPINDOWN THEOREM is true.

\vskip 0.2cm

\noindent
--we consider firstly evolution of spin period of a NS on the Accretor stage
and predict that its period  $ \geq 5\cdot 10^2 $ sec and INSs can be observed
as pulsating X--ray sources.

\vskip 0.2cm

\noindent
--we modeled accretion  onto  an INS
from the interstellar medium in the  case  of
spherical symmetry for different values of  the  magnetic  field
strength, ambient gas density and NS's mass.
The  periodic  sources  with  $P$  from
several minutes to several months can appear.

\vskip 0.2cm

\noindent
--we consider new idea of stochastic acceleration of  very old NSs due to
accretion of turbulizated ISM.

\vskip 0.2cm

\noindent
--last point means that INSs can be spin up and spin down
with equal probability.

\vskip 0.2cm

\noindent
--using the observed period changes for four systems:
Vela X--1, GX 301--2, Her X--1 and Cen X--3 we determined {\it D}, the
'diffusion coefficient',--parameter from the Fokker--Planck equation.

\vskip 0.2cm

\noindent
--using strong dependence of {\it D} on the velocity for Vela X--1 and GX 301--2,
systems accreting from a stellar wind, we determined the stellar wind velocity.
For different assumptions for a turbulent velocity we obtained
 $V=(660-1440)\, km\cdot s ^{-1}$.

\vskip 0.2cm

\noindent
--we also determined the specific characteristic time scales for
the 'diffusion processes' in X-ray pulsars. It is of order of 200 sec
for wind-fed pulsars and 1000-10000 sec for the disk accreting systems.

\end{abstract}

\clearpage

\section{Introduction}

\vskip 0.1cm

 Isolated neutron stars (INSs) which are not observable as
radiopulsars
became recently to attract much interest. Treves $\&$ Colpi (1991)
 suggested that INSs accreting from interstellar  medium (ISM) can be
observed in UV and X--rays by {\it ROSAT} and several of them were observed.

\vskip 0.1cm

For main--sequence stars there are two the most important parameters which
are needed for description of their evolution: the mass and the age.
For NSs there are also two important parameters: the period and
the gravimagnetic parameter,$y\,$.

 The gravimagnetic parameter was firstly introduced by
Davis $\&$ Pringle (1981) as:

 \begin{equation}
 y=\frac{\dot M}{\mu^2}
 \end{equation}

\vskip 0.1cm

 There are four possible states of a NS in low density plasma: E (Ejector),
P (Propeller), A (Accretor) and G (georotator), depending on relations
between four characteristic radii: $R_l$, radius of the light cylinder,
$R_{st}$, stop radius , $R_G=(\frac {2\, G\, M}{v_\infty ^2})$--radius of
gravitational capture and $R_c=(\frac {G\, M}{\omega ^2})^{1/3}$--radius of corotation.

 As the result we have two critical periods: $P_E $ and $P_A$, separating
different stages of NS. If $P<P_E$  we have Ejector ,
if $P_E<P<P_A$ we have the Propeller stage and if $P>P_A$ and
$R_{st}<R_G$ --the Accretor
stage. In some cases it may be $P>P_A$ but $R_{st}>R_G$ and
accretion is not possible
because Geo--like magnitosphere is formed.

\vskip 0.1cm

 Usially a track of a NS can be represented as a monotonic spin--down,
consequantly passing through three stages:

$$
 E\longrightarrow P\longrightarrow A
$$
$$
 E\longrightarrow P\longrightarrow G
$$

 The stage of Georotator can take place if:

 \begin{equation}
 y<y_G=\frac {v_{\infty} ^7}{(2\, G\, M)^4}
 \end{equation}

\vskip 0.1cm

 On the Fig.1 we present  examples of the NS evolutional tracks.

 In this poster we show some results, concerning periods of NSs, their
evolution, determination of the stellar wind velocities in X-ray pulsars
and INSs. Mostly we are based on three our articles (Popov 1994,
Lipunov and Popov 1995a,b)

\clearpage

\section{The Spindown theorem.}

 The equation of evolution of INSs can be represented in the form of the
spin--down equation (Lipunov 1982):

 \begin{equation}
 \frac {d\, I\omega}{dt}=-\frac {k_t\mu ^2}{R_t^3}
 \end{equation}
where $k_t$--dimensionless factor, $R_t$--the scale of the interaction between
the magnetic field of the NS and surrounding plasma.

\vskip 0.1cm

For the next consideration we proposed some kind of astrophysical theorem
from the general point of view.

\vskip 0.2cm

{\it THE SPINDOWN THEOREM.}

\vskip 0.1cm

 \it
 The duration of the Propeller stage under all other constant conditions
is smaller than duration of the Ejector stage:

$$
 t_P\le t_E, \qquad (SDT)
$$

\vskip 0.1cm

 Arguments.
 \rm

\vskip 0.1cm

  I). For the most investigated propeller regimes
the spin--down coefficient, $k_t$,
slowly depends on the spin frequency, $\omega $, and as result the duration
time of the Propeller stage is determined by the initial (not final, as for
Ejector) period of the NS!

II). The interaction between the magnetic field of the NS and accreting plasma in each
case is more effective
than the interaction between the magnetic field and vacuum
(magnito-dipole spin--down). Realy,  we have the relation:

 \begin{equation}
 \beta=\frac {(\frac {d\, I\omega }{dt})_E}{(\frac {d\, I\omega }{dt})_P}=
 \frac 2{3k_t}\cdot \left(\frac {R_A}{R_l}\right)^3.
 \end{equation}

 Because $(R_A/R_l)^3$ is a very small parameter we can expect that $\beta <1$,
for a very large range of $k_t$.

\vskip 0.1cm

 {\it Conclusion}.

\vskip 0.1cm

If 1).and 2). are true then the duration of the Propeller stage
is less or equal than the duration of the Ejector stage.

\vskip 0.1cm

 If we believe in the SPINDOWN THEOREM we conclude that after $10^{7-8} $ yrs
the INS in real interstellar medium comes to the Accretor regime.

\vskip 0.1cm

 We have {\it strong observational argument} for the SPINDOWN THEOREM:
the X--ray pulsar
X Per is a very slowly rotating NS (period p=835 s) associated with young
massive companion (age of companion $\approx 10^7 yrs$) and has a very low
accretion rate (about $\approx10^{13-14} \, g/s$ approching to interstellar
medium). {\it We can not understand the existance of this X-ray pulsar if the
propeller mechanism is not effective!}

\section{Stochastic acceleration of INSs.}

 Now we will discuss the situation with very old INSs and their periods.
 Let us consider the fluctuations of the period of the NS. Suppose that the
change in the angular velocity of the accreting star occurs under the action
 of a random torque:

\begin{equation}
 \frac {d\omega}{dt}=-\frac {k_t\mu ^2}{IR_c^3}+\Phi
\end{equation}
here $\Phi$ is the fluctuating torque due to accretion from turbulizated
ISM. We can introduce the scalar potential $V$
(see Lipunov 1992):

\begin{equation}
 \frac {k_t\mu ^2}{IR_c^3}=\nabla_{\omega}V(\omega)
\end{equation}

 The distribution function among frequency, $f(\omega)$, can be recieved as
a result of stationary solution corresponding the Fokker--Planck equation:

\begin{equation}
f(\omega)=N\cdot \exp {(-V(\omega)/D)}
\end{equation}
here $D$--the diffusion coefficient. For our case we have:

\begin{equation}
V(\omega)=\frac {k_t\mu ^2}{3GMI}\mid \omega \mid ^3
\end{equation}

and

\begin{equation}
D\approx \frac12\left(\frac {\dot{M} v_tR_t}I\right)^2\frac {R_G}v
\end{equation}
where $v_t$ and $R_t$--velocity and length of the turbulence. We must put
$R_t\approx R_G,$ and after some calculations we obtain:

\begin{equation}
D\approx 7.6\cdot 10^{-18}\rho_{-24}^2v_6^{-13}v_{t_6}^2I_{45}^{-2}
\left(\frac M{M_\odot}\right)^7
\end{equation}

 Assuming $V(\omega)\approx D$ we obtain the estimate of the possible period
of INS due to stochastic acceleration:

$$
P=2\pi\left(\frac {k_t\mu^2}{3GMID}\right)^{1/3}\approx
$$

\begin{equation}
\approx 5\cdot 10^2 k_t^{1/3}\mu_{30}^{2/3}I_{45}^{1/3}
\rho_{-24}^{-2/3}v_6^{13/3}v_{t_6}^{-2/3}\left(\frac M
{M_\odot }\right)^{-8/3} \, s.
\end{equation}

 In principle this estimates show that {\it real period of INS can be about
several hours or days}. But due to the very strong dependence on the velocity,$v$,
and the turbulent velocity, $v_t$, at the scale about $R_G$ we cannot give more
precise results. We only illustrate that {\it the old INSs can be accelarated by
stochastic angular momentum from the turbulisated ISM.
 In this case we can see with equal probability spin--up and spin--down
INSs!}

\vskip 0.1cm

 Spin--up and spin--down time can be estimated as:

$$
t_{su}\approx t_{sd}\approx \frac {I\omega}{\dot{M} v_tR_G}=
\frac {Iv_{\infty}^5}{(2GM)^3\rho_{\infty}v_tp}\approx
$$

\begin{equation}
 \approx 20\, yrs\, I_{45}v_6^5\rho_{-24}^{-1}\left(\frac p{10^5\, s}\right)^{-1}
 \left(\frac {v_t}{10^6\, cm/s}\right)^{-1}
\end{equation}

 Realy $v_t$ is a small fraction of the sound velocity, because in opposite
case the energy of turbulence will dissipate in the form of shock waves.
From observations it is well known that turbulence has a power spectrum:

\begin{equation}
v_t\sim l^{\alpha},
\end{equation}
 (here $l$ is the scale of the turbulent torque with
turbulent velocity $v_t$), but exact value of $\alpha$ is unknown. Very often
the turbulence spectra is represented in the form of the Kolmogorov spectrum:

\begin{equation}
 v_t^3=\epsilon l,
\end{equation}

where $\epsilon $ is a constant which characterises energy transfer. If we
want to calculate $v_t$ on the scale of $R_G$ we must know $v_t$ on the
characteristic scale. Turbulence is observed on scales $ 10^8<l<3\cdot 10^ {20}\, cm$
and for $l=(50-150)\, pc$ $v_t$ is of order of $10^6\, cm/s$ (Ruzmaikin et al 1988).
we also use another value: $v_t=10^5(\frac l{10^{18}cm})^{1/3}$ (Kaplan $\&$
Pikelner 1979). so for $R_G=4\cdot 10^{14}\, cm $ we have:
$v_t\approx10^4-10^5\, cm/s$.
In this case $t_{su}$ and $t_{sd}$ are of order of 200-2000 yrs.
In dense clouds ($\rho_{-24}\approx 100$) $v_t\approx 3\cdot 10^3-10^4\,
 cm/s$ (Canuto $\&$ Battaglia 1985) and $t_{su}\approx t_{sd}\approx 20-60\, yrs$
 and NS can be observed in principle.

\subsection{Accretion onto OINSs and periodic sources}

 We modeled accretion  onto an INS
 from the interstellar medium in the  case  of
spherical symmetry for different values of  the  magnetic  field
strength, ambient gas density and NS's mass. We  tried  to  verify
the idea that if the radius of corotation, $R_{co}$,
is less than the Alfven radius, $R_A$ ,
the shell will form around  the  INS  and   $R_A$
will decrease to $R_{co}$   and the periodic X-ray source  will  appear
(see Colpi et al., 1993).

\vskip 0.1cm

Dependence of  $R_A$  from $t$ in our model coincides  well  with  the
analytic formula from Colpi et al.(1993).
{\it The  periodic  sources  with  $P$  from
several minutes to several months can appear}
(see Popov 1994).

\section{The 'Diffusion coefficients' and
the stellar wind velocities for X--ray binaries.}

 Here we shall try to estimate the diffusion coefficients for selected
systems (see section "Stochastic acceleration of INSs") and stellar
wind velocities (see details in Lipunov $\&$ Popov, 1995a,b).
 At first we shall estimate $D$.
All variables, except $ \Delta t $ , are known (in principle). Their
values taken from Lipunov (1992) are shown in table 1.

 Characteristic time $ \Delta t $ for the wind--accreting systems can be
determined from the equation:

 \begin{equation}
 \Delta t \approx 1.7\cdot 10^{4} \alpha^{-2}10^{2(A+8.5)}L_{37}^
 {-\frac {12}{7}}\mu_{30}^{-\frac {4}{7}} sec,
 \end{equation}
where $ \alpha $ -- a fraction of the specific angular momentum of
the Kepler orbit at the magnitospheric radius, $A$--noise level (see table 2)
(de Kool $\&$ Anzer 1993).

\vskip 0.1cm

 So we can write equation for $D$ in the form:

 \begin{equation}
 {D}=5.55\cdot 10^{-19}\mu _{30}^2\omega ^3
 \gamma _6^{-1}I_{45}^{-1}\left(\frac M{1.5M_{\odot}}\right)^{-1}\, s^{-3} \, .
 \end{equation}

Values of $D$ for four systems are shown in table 3.

For characteristic time in the frequency space we can write:
\begin{equation}
 t_{char}=\omega_{char}/{\dot{\omega}} =\frac {Dp^4}{4\pi ^2 {\dot{p}^2}}
\end{equation}

We can give a physical interpretation for $t_{char}$ for wind-fed pulsars
as a characteristic time of the momentum transfer:

\begin{equation}
  \frac {R_G}{v_{sw}}=400\left(\frac M{1.5M_\odot}\right)
  \left(\frac {v_{sw}}{10^8cm/s}\right)^{-3}\qquad  sec
\end{equation}

These characteristic time scales are also shown in table 3.

 Now for $D$ we can write:

 \begin{equation}
 {D}=4.38\cdot 10^{-23}{\dot{M}} _{16}^2\eta ^2 T_4 \mu ^{-1}v_{8}^{-7}
 I_{45}^{-2}\left(\frac M{1.5\, {M_\odot} }\right)\, s^{-3}\, .
 \end{equation}

 As we see {\it there is a strong dependence of $D$ on $v.$} So we can
evaluate $v$ (in this case it is the stellar wind velocity, $v_{sw}$):

 \begin{equation}
  v_{sw}=1700\cdot D_{-24}^{-1/7}{\dot{M}} _{16}^{2/7}\eta ^{2/7}T_4^{1/7}
 \mu ^{-1/7}
  I_{45}^{-2/7}\left(\frac M{1.5\, {M_\odot} }\right)^{2/7} \, km/s.
 \end{equation}

 Values of $v_{sw}$ are shown in table 3.

\clearpage
\begin{table*}
\caption[]{}
\begin{tabular}{|l||c|c|c|c|c|c|}
           & p, sec & p$_{orb}$, sec & L, erg/sec        & $\mu /10^{30}\, Gs\cdot cm^3$ & t$_{su\, observ}$, yrs & t$_{su\, min}$, yrs\\
 Vela X--1 & $283$  & $7.7\cdot 10^5$& $1.5\cdot 10^{36}$& $3$  & $3000$& $3000$\\
 GX 301--2 & $696$  & $3.6\cdot 10^6$& $10^{37}$         & $120$& $>100$& $100$ \\
 Her X--1  & $1.24$ & $1.5\cdot 10^5$& $10^{37}$         & $0.6$& $3\cdot 10^5$& $8000$\\
 Cen X--3  & $4.84$ & $1.8\cdot 10^5$& $5\cdot 10^{37}  $& $5.7$& $3400$& $600$\\
\\
\end{tabular}
\end{table*}

\begin{table}
\caption[]{}
\begin{tabular}{|l||c|c|}
           &  A   & L$_{max}$  \\
Vela X--1  &$-9.1$& $10^{36.8}$\\
GX 301--2  &$-8.5$& $10^{37}  $\\
\\
\end{tabular}
\end{table}

\begin{table*}
\caption[]{}
\begin{tabular}{|l|c|c|c|c|c|c|}
          &$\Delta t,\, sec$&$\gamma  $     &$D,\, sec^{-3}$    &$v_{sw},\, km/s,\,
 $&$v_{sw},\, km/s,\, $&$t_{char},\, sec$\\
          &  &  &  & $(v_t=0.1\cdot a_s)$ & $(v_t=a_s)$ &  \\
Vela X--1 &$1.5\cdot 10^4$  &$6.3\cdot 10^6$&$8.7\cdot 10^{-24}$&$848$ &$1442$ &$200$\\
GX 301--2 &$1.1\cdot 10^3$  &$2.9\cdot 10^6$&$2  \cdot 10^{-21}$&$656$ &$1120$ &$245$\\
Her X--1  &$4\cdot 10^3  $  &$6.0\cdot 10^7$&$4  \cdot 10^{-19}$&---   &---    &$960$\\
Cen X--3  &$2\cdot 10^4  $  &$9.5\cdot 10^5$&$4.1\cdot 10^{-17}$&---   &---    &$8500$\\
\\
\end{tabular}
\end{table*}

\clearpage

\section{Aknowledgements}

	This work was supported by RFFI and INTAS.
P.S.B. also thanks the ISF (for travel support) and  ISSEP.

 More information on the topic can be found at:
http://xray.msu.su/~polar/.
Your comments you can send to:polar@xray.sai.msu.su or ps@sai.msu.su.

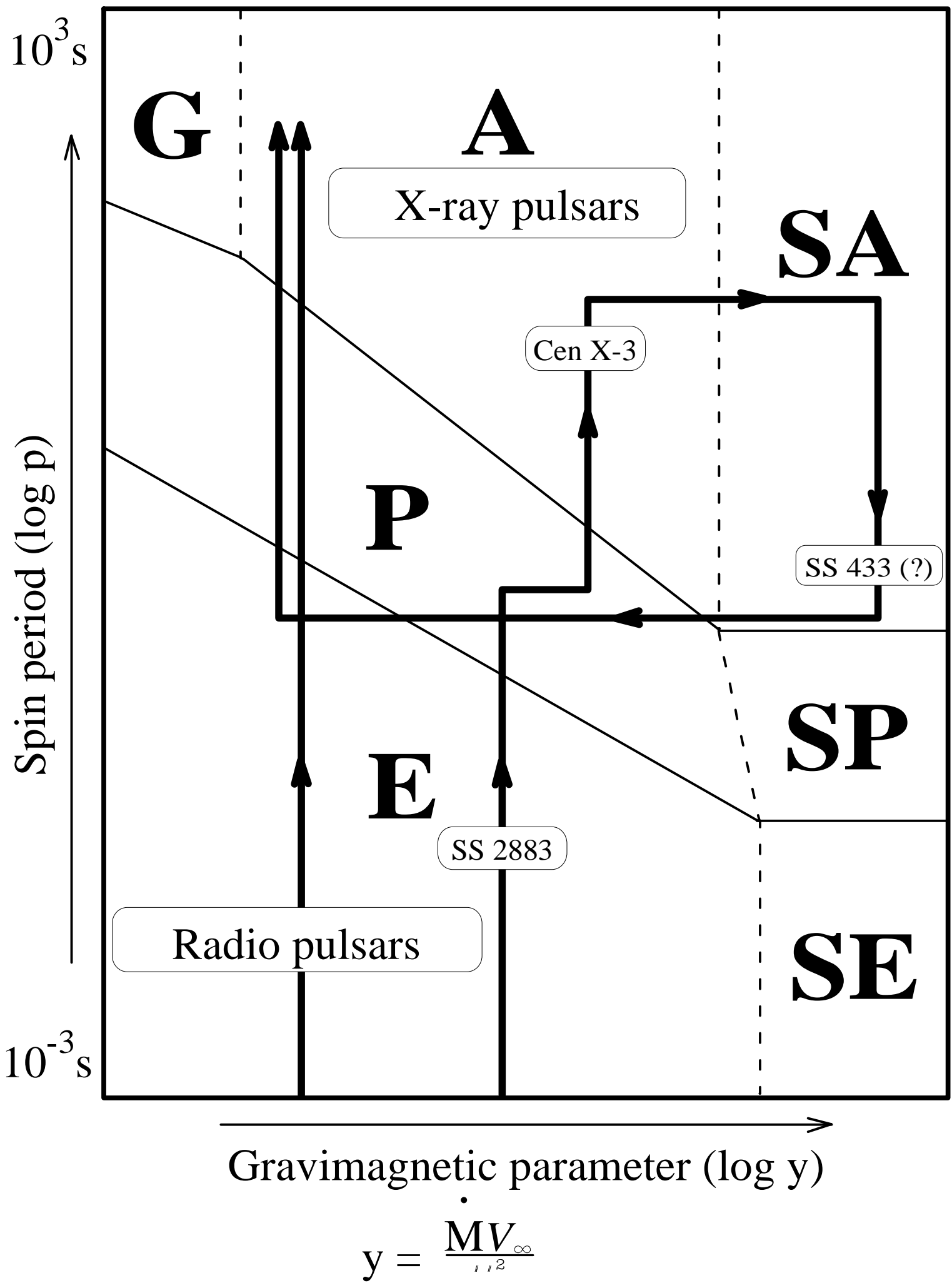
\end{document}